\documentclass[prd,aps,showpacs,,nofootinbib,superscriptaddress,preprintnumbers,epsf,psf]{revtex4}
\usepackage[english]{babel}
\usepackage{pstricks}
\usepackage[dvips]{graphicx}
\usepackage{epsf}
\usepackage{amsmath}
\usepackage{amsfonts}
\usepackage{amssymb}
\usepackage{epsf}
\usepackage[dvips]{graphicx}
\usepackage{dcolumn}
\usepackage{bm}
\everymath{\displaystyle}
\begin{document}

\title{Helicity separation in Heavy-Ion Collisions }

\author{\firstname{Mircea}~\surname{Baznat}}
\email{baznat@theor.jinr.ru}
\author{\firstname{Konstantin}~\surname{Gudima}}
\email{gudima@theor.jinr.ru} \affiliation{Joint Institute for Nuclear Research, 141980 Dubna (Moscow region),
Russia} \affiliation{Institute of Applied Physics, Academy of Sciences
of Moldova, MD-2028 Kishinev, Moldova }
\author{\firstname{Alexander}~\surname{Sorin}}
\email{sorin@theor.jinr.ru}
\author{\firstname{Oleg}~\surname{Teryaev}}
\email{teryaev@theor.jinr.ru} \affiliation{Joint Institute for Nuclear Research, 141980 Dubna (Moscow region), Russia}
\affiliation{Dubna International University, Dubna (Moscow region) 141980, Russia}

\date{\today}

\begin {abstract}
We study the P-odd effects related to the vorticity of
the medium formed in noncentral heavy ion collisions.
Using the kinetic Quark-Gluon Strings Model we perform
the numerical simulations of the vorticity and hydrodynamical
helicity for the various atomic numbers,
energies and centralities. We observed the vortical structures
 typically
occupying the relatively
small fraction of the fireball volume. In the course of numerical
simulations the noticeable hydrodanamical helicity was observed
manifesting the specific mirror behaviour with respect to the
reaction plane. The effect is maximal at the NICA and FAIR energy range.


\end{abstract}


\maketitle

\section{Introduction}

The local violation \cite{Fukushima:2008xe} of discrete symmetries
in strongly interacting QCD matter is now under intensive
theoretical and experimental investigations. The renowned Chiral
Magnetic Effect (CME) uses the (C)P-violating (electro)magnetic
field emerging in heavy ion collisions in order to probe the
(C)P-odd effects in QCD matter.

There is an interesting counterpart of this effect, Chiral
Vortical Effect (CVE)\cite{Kharzeev:2007tn} due to coupling to
P-odd medium vorticity. In its original form
\cite{Kharzeev:2007tn} this effect leads to the appearance of the
same electromagnetic current as CME. Its straightforward
generalization was proposed later resulting in generation of all
conserved-charge currents \cite{Rogachevsky:2010ys}, in particular
baryonic ones (especially important when there is CME cancellation
between three massless flavors \cite{Kharzeev:2010gr}), and
polarization of hyperons \cite{Rogachevsky:2010ys,Gao:2012ix}. Let
us also mention quite recent very interesting theoretical
development \cite{Landsteiner:2011iq}, where a remarkable relation
to the gravitational anomalies was discovered.

The key problem is whether the flows developing in heavy ion collisions indeed possess the vorticity?!
This is especially interesting as vorticity is universal phenomenon  manifested at very different scales of
macro and micro physics. One can observe it in spiral galaxies, cyclones and typhoons, semiconductors,
chemical reactions, biological systems, quantum field theories, etc. It would be indeed very important to push
this concept furtherto the internal structure of QCD matter.

The noncentral heavy ion collisions could naturally generate a rotation
(global or local, both related to vorticity)
with an angular velocity normal to the reaction plane,
which is their generic qualitative feature.
However finding proper quantitative characteristics of these phenomena requires special investigation
\cite{Becattini:2009wh,Betz:2007kg}.
In this paper we address this problem using
the Quark-Gluon Strings Model (QGSM) and observe the clear signs and manifestations of vortical
and helical structures in QCD matter formed in noncentral heavy ion collisions.
In particular we observed the novel effect of the hydrodinamical helicity separation.

\section{Modelling velocity, vorticity and helicity in kinetic model}

One of the first models designed to describe the dynamics of energetic
heavy-ion collisions was the intra-nuclear cascade model developed in Dubna
\cite{toneev83} which is based on the Monte-Carlo solution of a set of the
Boltzmann-Uehling-Uhlenbeck relativistic kinetic equations with
the collision terms, including cascade-cascade
interactions. For particle energies below 1~GeV it is sufficient to
consider only nucleons, pions and deltas. The model includes a proper
description of pion and baryon dynamics for particle production and
absorption processes.
In the original version the nuclear potential is treated dynamically, i.e.,
for the initial state it is determined using the Thomas-Fermi approximation,
but later on its depth is changed according to the number of knocked-out
nucleons. This allows one to account for nuclear binding.
The Pauli principle is implemented by introducing a Fermi distribution
of nucleon momenta as well as a Pauli blocking factors for scattered
nucleons.

At energies higher than about 10~GeV, the Quark-Gluon String Model
(QGSM) is used to describe elementary hadron collisions
\cite{toneev90,amelin91}. This model is based on the 1/N$_c$
expansion of the amplitude for binary processes where N$_c$ is the
number of quark colours. Different terms of the 1/N$_c$ expansion
correspond to different diagrams which are classified according to
their topological properties. Every diagram defines how many
strings are created  in a hadronic collision and which
quark-antiquark or quark-diquark pairs form these strings. The
relative contributions of different diagrams can be estimated
within Regge theory, and all QGSM parameters for hadron-hadron
collisions were fixed from the analysis of experimental data. The
break-up of strings via creation of quark-antiquark and
diquark-antidiquark pairs is described by the Field-Feynman method
\cite{field78}, using phenomenological functions for the
fragmentation of quarks, antiquarks and diquarks into hadrons. The
modified non-Markovian relativistic kinetic equation, having a
structure close to the Boltzmann-Uehling-Uhlenbeck kinetic
equation, but accounting for the finite formation time of newly
created hadrons, is used for simulations of relativistic nuclear
collisions. One should note that QGSM considers the two lowest
SU(3) multiplets in mesonic, baryonic and antibaryonic sectors, so
interactions between almost 70 hadron species are treated on the
same footing. This is a great advantage of this approach which is
important for the proper evaluation of the hadron abundances and
characteristics of the excited residual nuclei. The
energy extremes
were bridged by the QGSM extension downward in the beam energy \cite{amelin90}.

For investigation of dynamical formation of velocity $\vec v$ and
vorticity
$\vec \omega\ (\equiv rot\, \vec v)$
fields in relativistic heavy ion collision the coordinate space
was divided into {$50\times50\times 100$} cells of volume $dxdydz$
with $dx=dy=0.6fm$, $dz=0.6/\gamma$ fm, where $\gamma$ is the
gamma factor of equal velocity system of collision. In this
reference system the total momentum and total energy of the
produced particles were calculated in all cells for each of fixed
25 moments of time $t$ covering the interval of $10~fm/c$.

The results were averaged for about 10000 heavy ion collisions
with identical initial conditions. The spectator nucleons of
projectile or target ions, which at given time momentum do not
undergone any individual collision, were included in evaluation of
velocity. The velocity field in the given cell was defined by the
following double sum over the particles in the cell and over the
all simulated collisions:
\begin{equation}
\vec v~(x,y,z,t)\ = \ \frac{\sum_i\sum_j \vec P_{ij}}{\sum_i\sum_j  E_{ij}} \,  \label{v}
\end{equation}
where $\vec P_{ij}$ and $E_{ij}$ are the momentrum and energy of
particle $i$ in the collision $j$, respectively.
The vorticity was calculated using the discrete
partial derivatives.

We paid a special attention to the pseudoscalar characteristics of
the vorticity, that is the hydrodynamical helicity
$H \equiv \int d V  (\vec v \cdot \vec w)$
which is related to a number of interesting phenomena in
hydrodynamics
 and plasma physics, such as the turbulent dynamo and Lagrangian chaos. It might be compared the analog of
topological charge $Q \ = \ \int d^3 x J^0(x)$
where the current  $J^{\mu}\ = \ \epsilon^{\mu\nu\rho\gamma}u_{\nu} \partial_{\rho}u_{\gamma}$
(as usual, the four-velocity $u_{\nu}\ \equiv \ \gamma~(1,\, \vec v~)$)
contributes to the hydrodynamical anomaly
\cite{Son:2009tf} and the polarization of hyperons \cite{Rogachevsky:2010ys,Gao:2012ix}.
The calculation of the topological charge which is the correct relativistic generalization of the
hydrodynamical helicity leads to the extra factor $\gamma^2$ in the integrand.
Still as the helicity itself is a more traditional quantity, we use it for the numerical calculations.


\section{Results of the simulations: helicity separation effect}

The qualitative pictures of velocity and vorticity fields corresponding to
$Au+Au$ collisions at $\sqrt{s_{NN}}\ =\ 5 ~Gev$ with the impact parameter $8 ~fm$ equal to the (transverse)
radius of the nuclei are presented at figures 1-2.
\begin{figure}[h!]
\includegraphics[angle=-0,width=0.7\textwidth]{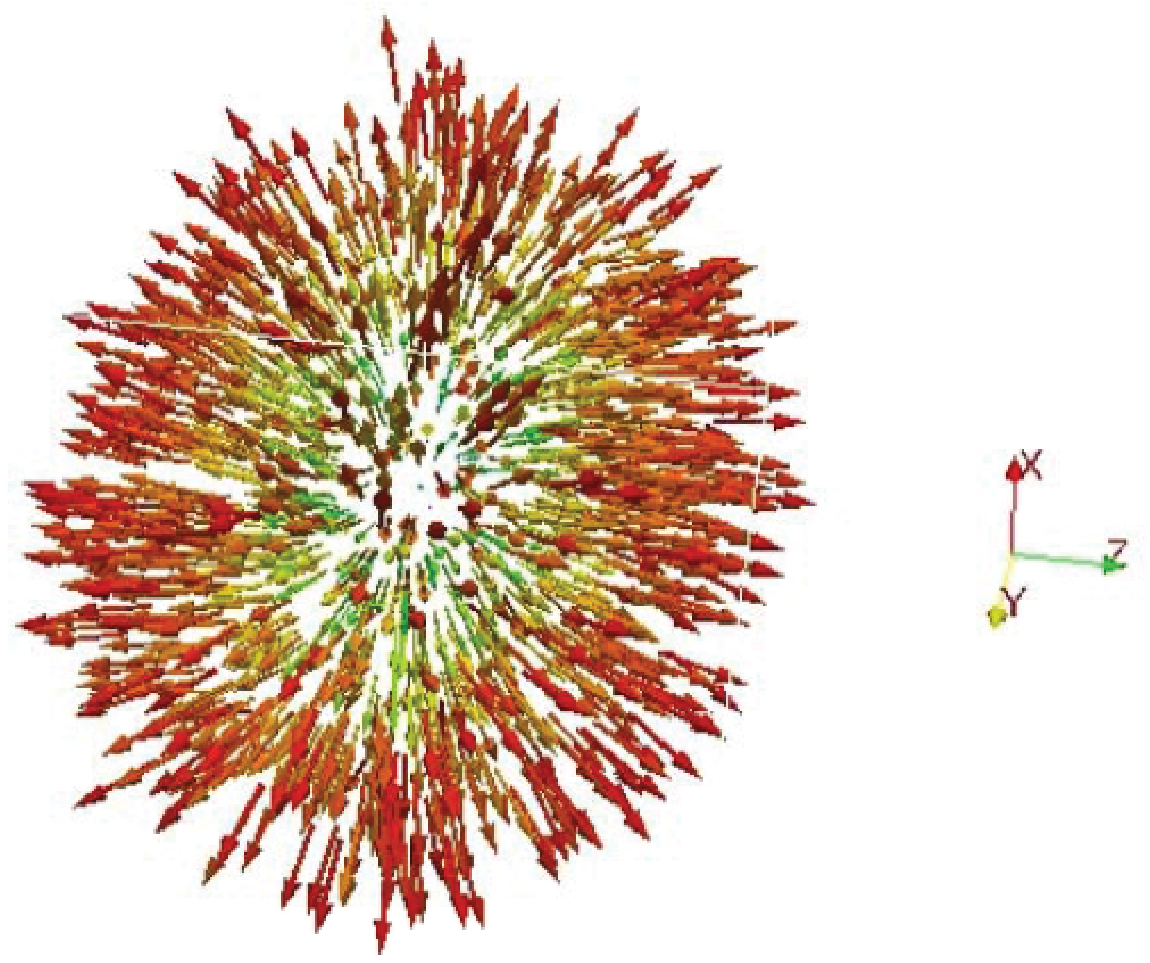}
\end{figure}
\begin{figure}[h!]
\includegraphics[angle=-0,width=0.5\textwidth]{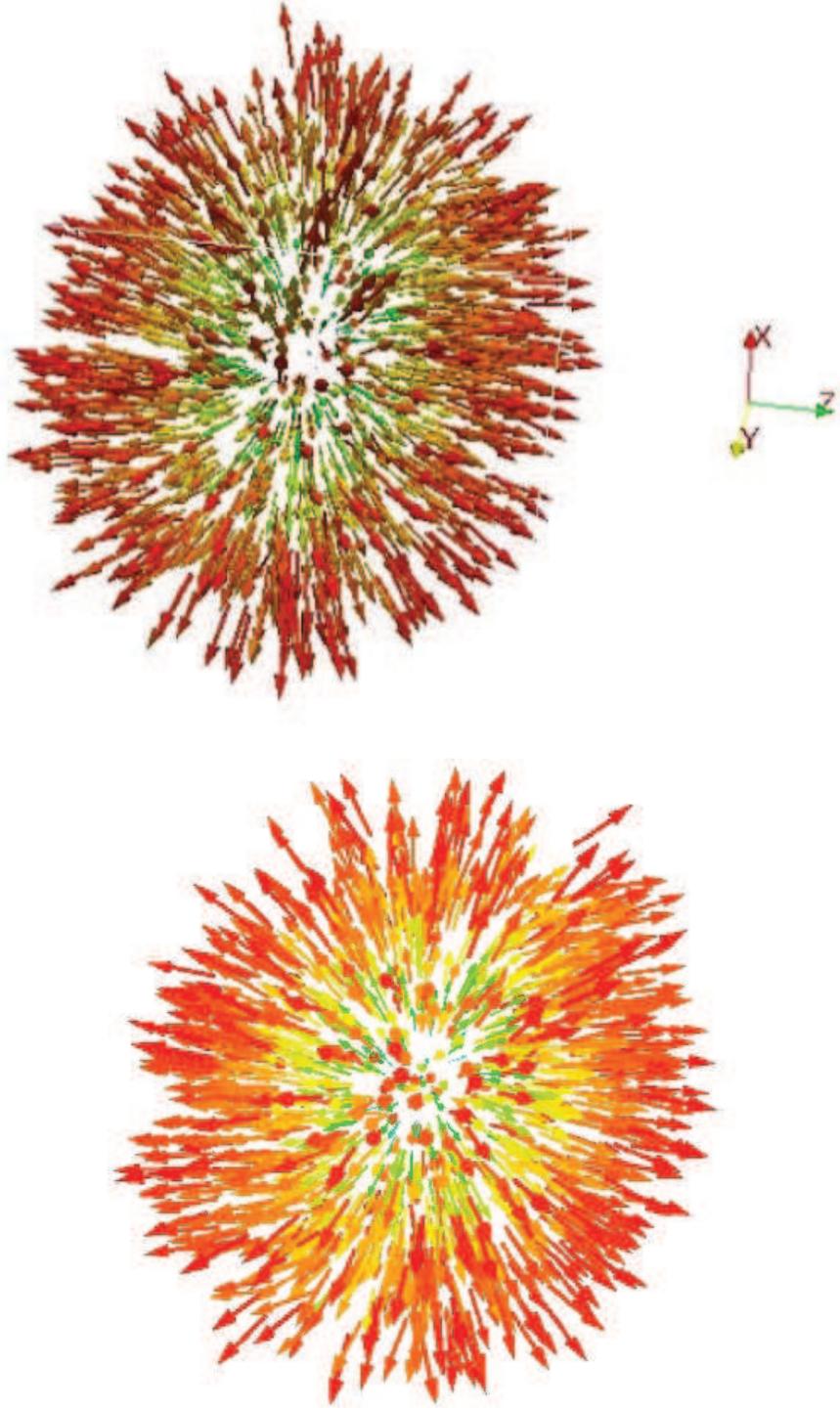}
  \caption{(color online) Three dimension image (top) and projection
  on plane $XY$ (bottom) of velocity field in Au+Au at $\sqrt{s_{NN}}\ =5~GeV$, $b=8~fm$ and $t=10~fm/c$}
  \label{Fig.1}
\end{figure}

Figure $1$ represents the three dimensional
distribution (top) of the velocity defined by the collision
participants and produced particles after $10~ fm/c$ of the
evolution, and its projection (bottom)
to the transverse $xy$-plane. The direction, length and color of the arrows represent
the direction and size of the velocity. We clearly see the
picture of a little bang when fastest particles (pions) are
occupy most distance positions from the collision origin.

Fig. 2 represents the similar distributions for the vorticity.
The vorticity is concentrated in the relatively thin $(2 \div 3
~fm)$ layer at the boundary of the participant  region. This
might be an analog of the vortex sheet expected when
Kelvin-Helmholtz instability develops \cite{Csernai:2011qq}. Let
us stress that in the case under consideration it emerges in the
kinetic approach which might be of some interest for the
microscopic description of turbulence.
\begin{figure}[h!]
\includegraphics[angle=-0,width=0.8\textwidth]{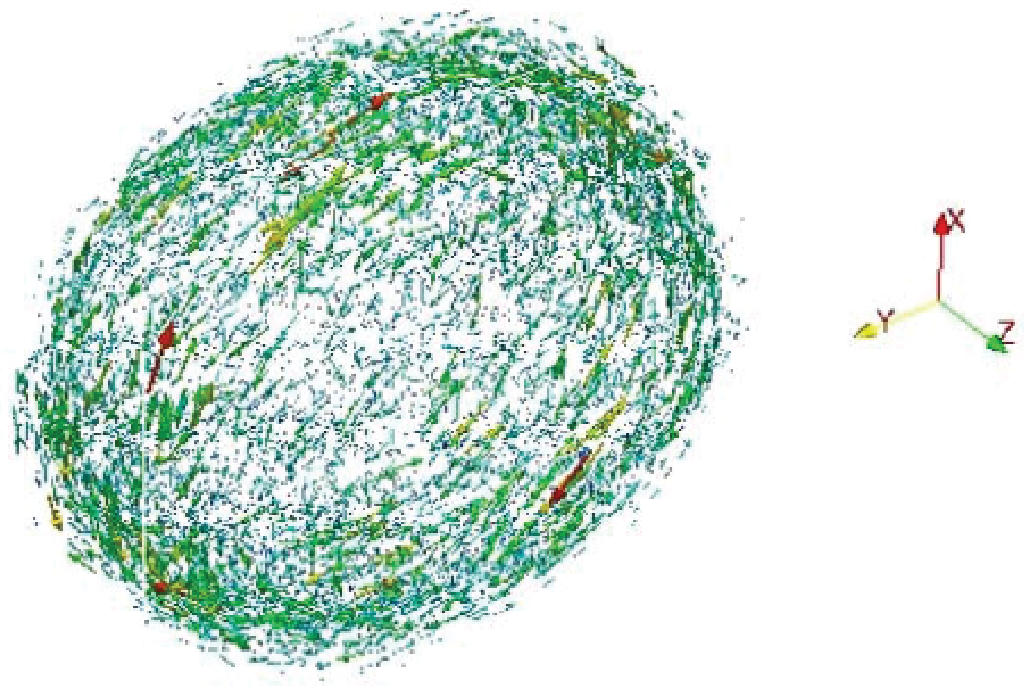}
\end{figure}
\begin{figure}[h!]
\includegraphics[angle=-0,width=0.6\textwidth]{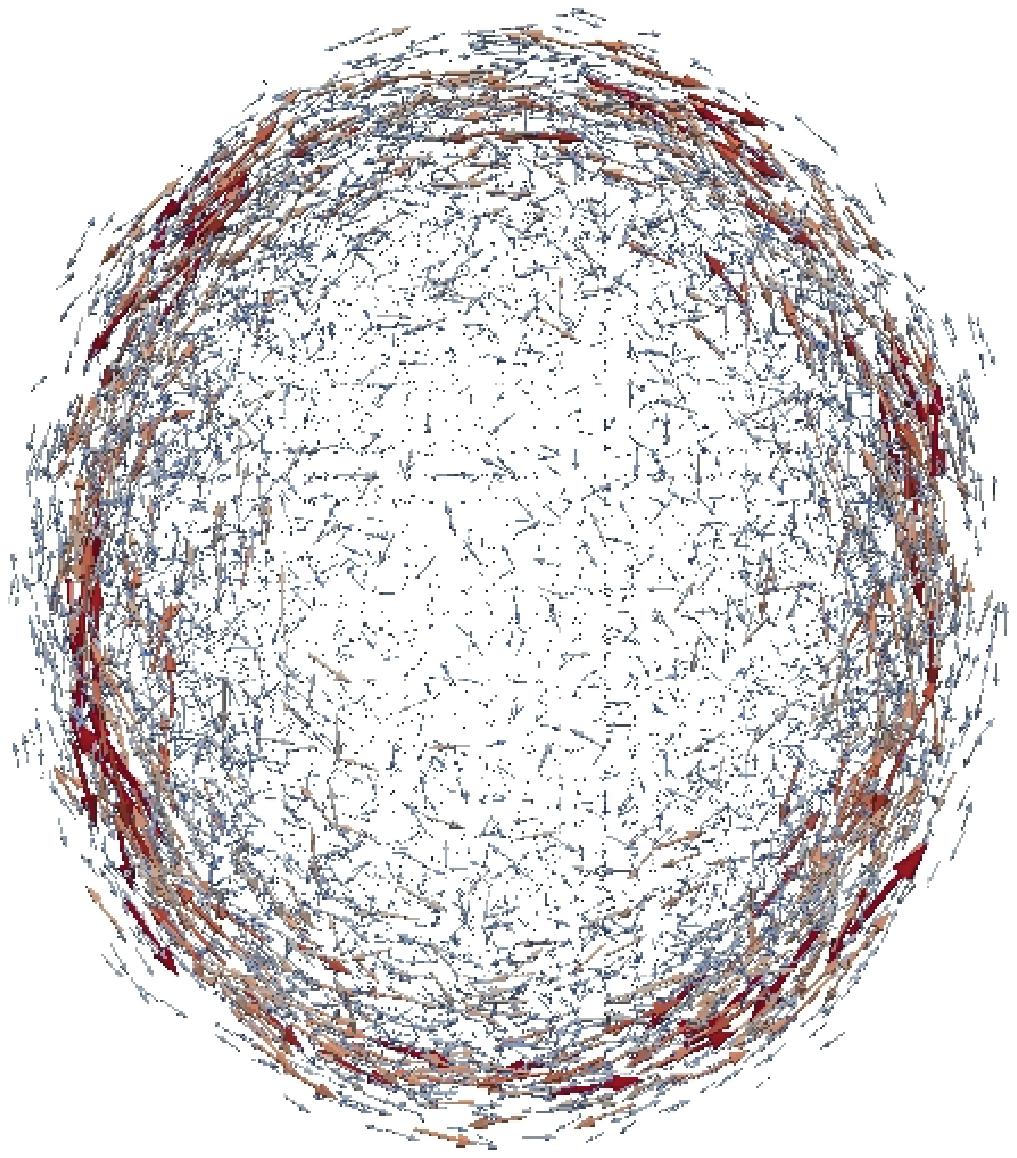}
  \caption{(color online) Three dimension image (top) and projection
  on plane $XY$ (bottom) of vorticity field in Au+Au at $\sqrt{s_{NN}}\ =5~GeV$,
 $b=8~fm$
  and $t=10~fm/c$}
  \label{Fig.2}
\end{figure}

For quantitative description of this phenomena we, as it was
already mentioned, use the hydrodynamic helicity whose patterns
are presented on Figures 3-6.
\begin{figure}[h!]
\includegraphics[angle=-0,width=0.6\textwidth]{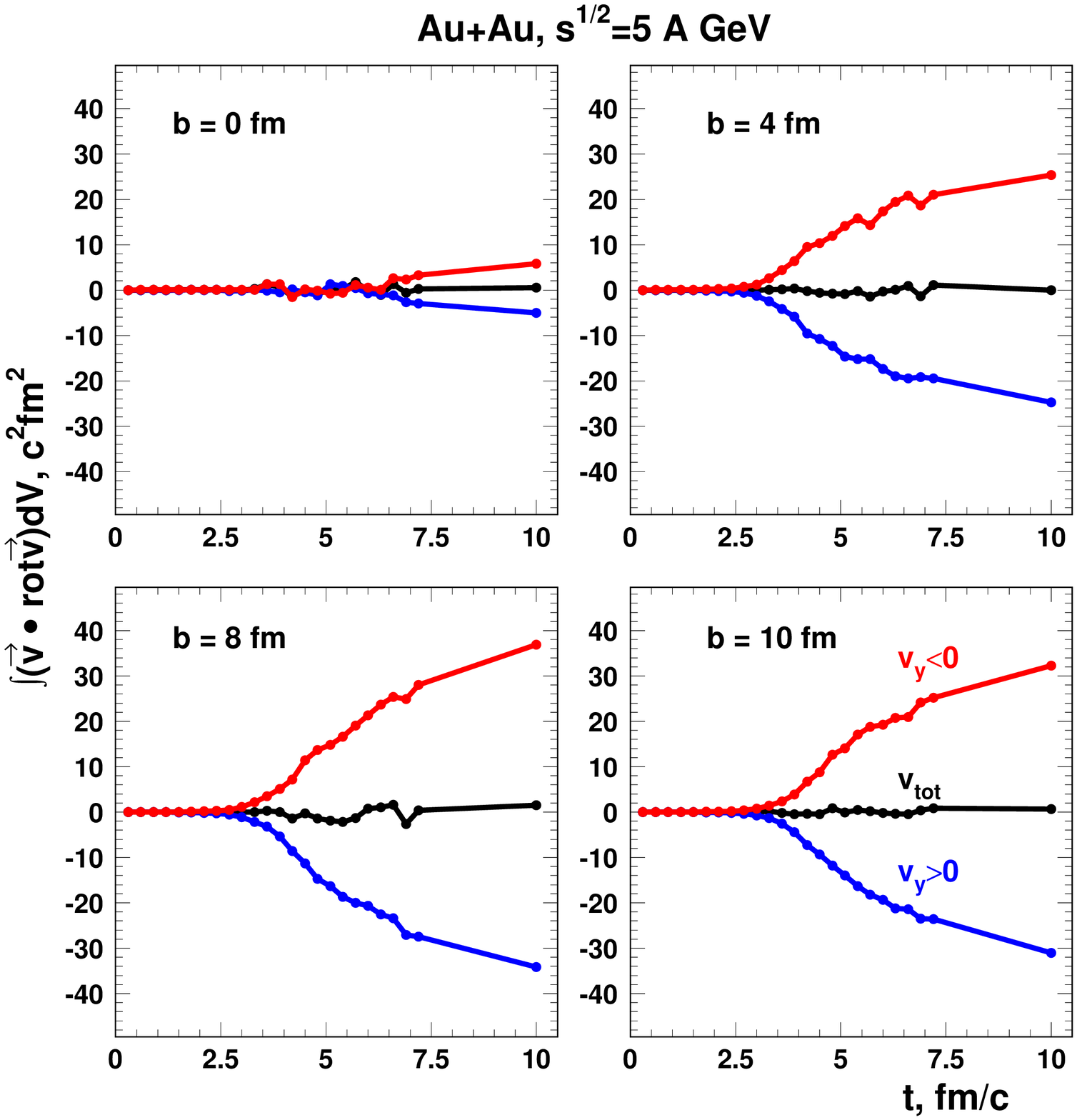}
  \caption{(color online) Time dependence of helicity at different impact parameters}
  \label{Fig.3}
\end{figure}

Fig. 3 represents the helicities in gold-gold collisions at
different impact parameters evaluated in different domains. One
can see that the helicity calculated with inclusion of the all
cells is zero (black line). For the cells with the definite sign
of the velocity components, which are orthogonal to the reaction
plane (which may be selected also  experimentally), the helicity
is nonzero and changes the sign for the different signs of these
components (red and blue lines, respectively). The effect is
growing with impact parameter and represents a sort of saturation
in time.
\begin{figure}[h!]
\includegraphics[angle=-0,width=0.6\textwidth]{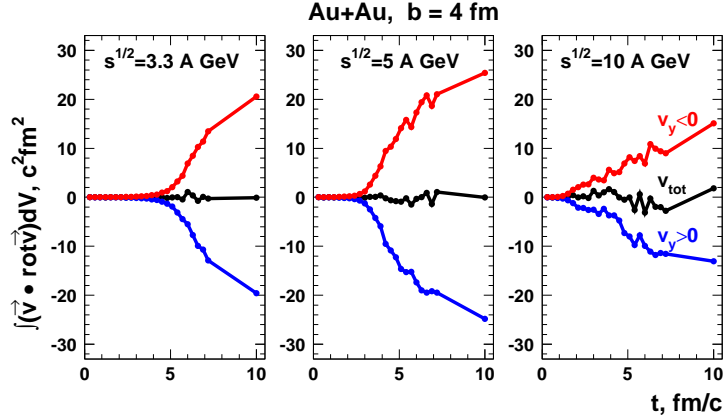}
  \caption{(color online) Time dependence of helicity at different energies}
  \label{Fig.4}
\end{figure}
This effect of helicity separation is one of our main results.
Let us stress that the calculation in the hybrid UrQMD model manifests very similar behavior \cite{Steinheimer}.
This is not surprising as the helicity is in fact generated at the hydrodynamical stage of the model, while the
transition from kinetic to hydrodynamical stage should be performed similarly to our Eq. (\ref{v}).

This effect might be qualitatively explained, if the perpendicular
components of velocities (which are selected to have different
signs) and the corresponding vorticities (assumed to have the same
signs) provide the dominant contribution to the scalar product in
the helicity definition. However, the numerical analysis showed
that the longitudinal components along the beam directions
(z-axis) provide even larger contribution to the helicity than
contributions from the transverse direction (y-axis). So, such
qualitative picture is oversimplified, but still provides a
correct sign convention for the helicity-separation effect.

 Fig. 4 represents the energy dependence of helicity which
shows that its maximal value is achieved around the NICA energy
range.

Fig. 5 shows how the effect is manifested in asymmetric collisions
of gold and argon ions in comparison to the gold-gold ones.

\begin{figure}[h!]
\includegraphics[angle=-0,width=0.6\textwidth]{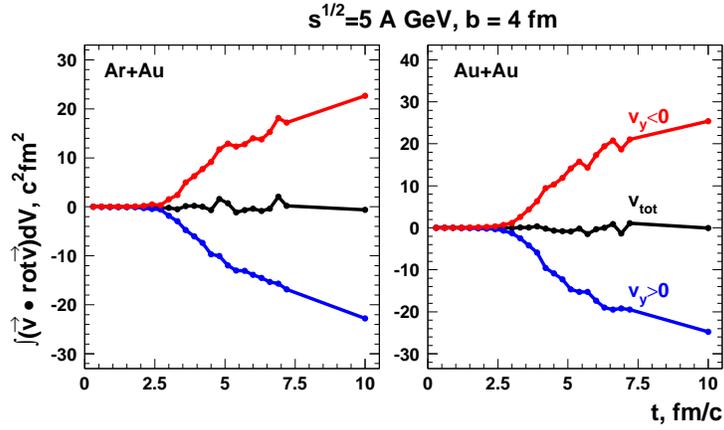}
  \caption{(color online) Time dependence of helicity in asymmetric collisions }
  \label{Fig.5}
\end{figure}
\begin{figure}[h!]
\includegraphics[angle=-0,width=0.6\textwidth]{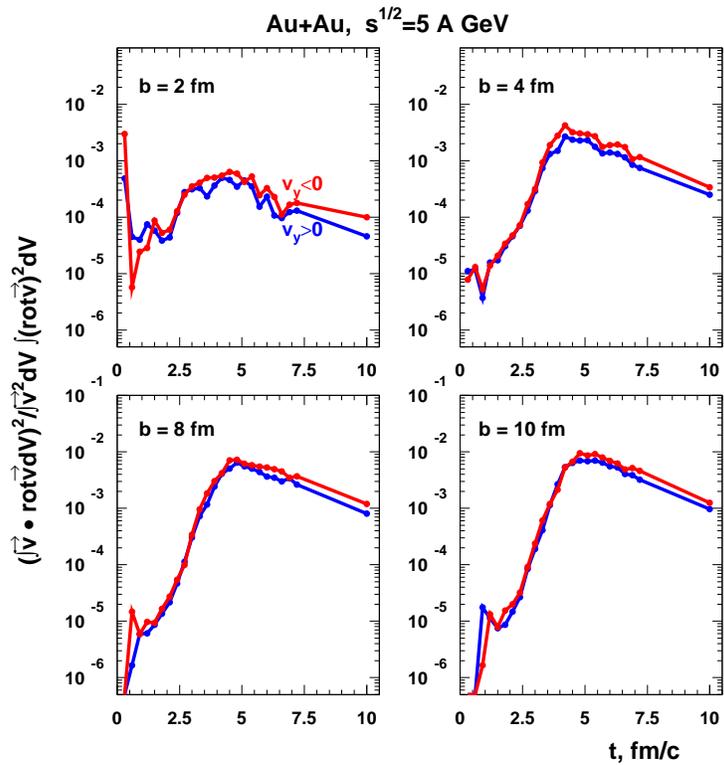}
  \caption{(color online) Time dependence of Cauchy-Schwarz bound for helicity }
  \label{Fig.6}
\end{figure}

Fig. 6 shows the dimensionless ratio bounded from above by
$1$ due to Cauchy-Schwarz inequality. This bound is saturated for
helical flows with vorticities parallel to velocities.
In the case of incompressible fluids the helicity of such flows is proportional
to the (non-relativistic) kinetic energy.
 The actual
values of this ratio show that the correlation between the
directions of the vorticity and the velocity is not large, but
non-negligible.

\section{Conclusions}

We investigated vorticity and hydrodynamical helicity in noncentral heavy-ion collisions in the framework of the kinetic quark-gluon string model. We have observed that the vorticity is predominantly localized in a relatively thin layer ($2\div 3~fm$) on the boundary between the participant and spectator nucleons. This might be qualitatively understood in the spirit of the core-corona type models \cite{Aichelin,St.& Bl.}.
Thus, the gradients of the velocities in the
region occupied by the participants are small due to the
compensation of momenta between the target and projectile
particles in the c.m. frame. As the result the vorticity is
substantial only in the thin transition layer between the participant
(i.e., core) and the spectator (i.e., corona) regions. We found
the novel effect of the helicity separation in heavy-ion
collisions when it has the different signs below and above of the
reaction plane. We have investigated its dependence on the type of nuclei and collision energy and observed
that it is maximal in the NICA energy range. We have also
calculated the degree of alignment of the velocity and vorticity
which is maximal for the Beltrami flows whose relativistic
generalization is currently under investigation
\cite{SorinTeryaev}.

{}~

{\bf Acknowledgements}

{}~

We are indebted to E.L. Bratkovskaya, K.A. Bugaev, L. Csernai,
V.D. Kekelidze,V.D. Toneev, V. Voronyuk, V.I. Zakharov, G.M.
Zinovjev and, especially, M. Bleicher, J. Steinheimer and  H.
St\"ocker for useful discussions and comments. This work was
supported in part by the Russian Foundation for Basic Research,
Grant No. 11-02-01538-a.


\end{document}